\documentclass[fleqn,usenatbib]{mnras}

\usepackage{newtxtext,newtxmath}

\usepackage[T1]{fontenc}

\DeclareRobustCommand{\VAN}[3]{#2}
\let\VANthebibliography\thebibliography
\def\thebibliography{\DeclareRobustCommand{\VAN}[3]{##3}\VANthebibliography}

\usepackage{graphicx}	
\usepackage{amsmath}	

\usepackage{graphicx}
\usepackage{amsmath}
\usepackage{threeparttable}
\usepackage{caption}
\usepackage{bibunits}
\usepackage{color}
\usepackage{adjustbox}
\usepackage{url}
\usepackage{soul}
\newcommand{\cii}{[C\,{\sc ii}]}

\newcommand{\hii}{H\,{\sc ii}}

\newcommand{\oiii}{[O\,{\sc iii}]}

\newcommand{\oiiil}{[O\,{\sc iii}] 88\,$\mu{\rm m}$}

\definecolor{referee}{RGB}{0,0,0}
\definecolor{referee2}{RGB}{0,0,0}

\title[ALMA observations on the $z\sim12$ galaxy candidate GHZ2/GLASS-z12]{Deep ALMA redshift search of a $z\sim12$ GLASS-JWST galaxy candidate }

\author[Bakx \& Zavala et al.]{Tom J. L. C. Bakx,$^{1,2}$\thanks{E-mail: bakx@a.phys.nagoya-u.ac.jp}
Jorge A. Zavala,$^{2}$\thanks{E-mail: jorge.zavala@nao.ac.jp}
Ikki Mitsuhashi,$^{3,2}$ Tommaso Treu,$^{4}$Adriano Fontana,$^{5}$ \newauthor Ken-ichi Tadaki,$^{2,6}$    Caitlin M. Casey,$^{7}$ Marco Castellano,$^{5}$ Karl Glazebrook,$^{8}$ Masato Hagimoto,$^{1}$  \newauthor Ryota Ikeda,$^{2,6}$ Tucker Jones,$^{9}$ Nicha Leethochawalit,$^{10,11,12}$ Charlotte Mason,$^{13,14}$ Takahiro Morishita,$^{15}$ \newauthor  Themiya Nanayakkara,$^{8}$ Laura Pentericci,$^{5}$ Guido Roberts-Borsani,$^{4}$ Paola Santini,$^{5}$ Stephen Serjeant,$^{16}$  \newauthor Yoichi Tamura,$^{1}$ Michele Trenti,$^{10,11}$ and Eros Vanzella$^{17}$\\
$^{1}$Division of Particle and Astrophysical Science, Graduate School of Science, Nagoya University, Aichi 464-8602, Japan\\
$^{2}$National Astronomical Observatory of Japan, 2-21-1, Osawa, Mitaka, Tokyo, Japan\\
$^{3}$Department of Astronomy, The University of Tokyo, 7-3-1 Hongo, Bunkyo, Tokyo 113-0033, Japan\\
$^{4}$Department of Physics and Astronomy, University of California, Los Angeles, 430 Portola Plaza, Los Angeles, CA 90095, USA\\
$^{5}$ INAF - Osservatorio Astronomico di Roma, via di Frascati 33, 00078 Monte Porzio Catone, Italy\\
$^{6}$Department of Astronomical Science, SOKENDAI (The Graduate University for Advanced Studies), Mitaka, Tokyo 181-8588, Japan\\
$^{7}$Department of Astronomy, The University of Texas at Austin, Austin, TX, USA\\
$^{8}$Centre for Astrophysics and Supercomputing, Swinburne University of Technology, PO Box 218, Hawthorn, VIC 3122, Australia\\
$^{9}$Department of Physics and Astronomy, University of California Davis, 1 Shields Avenue, Davis, CA 95616, USA\\
$^{10}$School of Physics, University of Melbourne, Parkville 3010, VIC, Australia\\
$^{11}$ARC Centre of Excellence for All Sky Astrophysics in 3 Dimensions (ASTRO 3D), Australia\\
$^{12}$National Astronomical Research Institute of Thailand (NARIT), Mae Rim, Chiang Mai, 50180, Thailand\\
$^{13}$Cosmic Dawn Center (DAWN)\\
$^{14}$Niels Bohr Institute, University of Copenhagen, Jagtvej 128, 2200 København N, Denmark\\
$^{15}$IPAC, California Institute of Technology, MC 314-6, 1200 E. California Boulevard, Pasadena, CA 91125, USA\\
$^{16}$School of Physical Sciences, The Open University, Milton Keynes MK7 6AA\\
$^{17}$INAF -- OAS, Osservatorio di Astrofisica e Scienza dello Spazio di Bologna, via Gobetti 93/3, I-40129 Bologna, Italy
}

\date{Accepted 2022 December 13. Received 2022 December 13; in original form 2022 August 30}

\pubyear{2022}

\begin{document}
\label{firstpage}
\pagerange{\pageref{firstpage}--\pageref{lastpage}}
\maketitle

\begin{abstract}
The James Webb Space Telescope (JWST) has discovered a surprising abundance of bright galaxy candidates in the very early Universe ($\leq 500$~Myrs after the Big Bang), calling into question current galaxy formation models. Spectroscopy is needed to confirm the primeval nature of these candidates, as well as to understand how the first galaxies form stars and grow. {\color{referee} Here we present deep spectroscopic and continuum ALMA observations towards GHZ2/GLASS-z12, one of the brightest and most robust candidates at $z > 10$ identified in the GLASS-JWST Early Release Science Program. We detect a $5.8 \sigma$ line, offset 0\farcs{}5 from the JWST position of GHZ2/GLASS-z12 that, associating it with the \oiiil{} transition, implies a spectroscopic redshift of $z = 12.117 \pm 0.001$. 
We verify the detection using extensive statistical tests.
The oxygen line luminosity places GHZ2/GLASS-z12 above the \oiii{}-SFR relation for metal-poor galaxies, implying an enhancement  of \oiii{} emission in this system  while the JWST-observed emission is likely a lower-metallicity region. The lack of dust emission seen by these observations is consistent with the blue UV slope observed by JWST, which suggest little dust attenuation in galaxies at this early epoch. Further observations will unambiguously confirm the redshift and shed light on the origins of the wide and offset line and physical properties of this early galaxy.} This work illustrates the synergy between JWST and ALMA and paves the way for future spectroscopic surveys of $z > 10$ galaxy candidates.
\end{abstract}




\section{Introduction}
The James Webb Space Telescope (JWST) recently opened a new window to the Universe with unprecedented sensitivity and angular resolution at near-infrared (NIR) wavelengths. The public release of the JWST Early Release Observations (ERO) and the Director’s Discretionary Early Release Science Programs (DD-ERS) have unlocked new searches for the faintest, rarest, and most distant galaxies ever found. 
Notably, the high sensitivity of the NIRCam instrument (\citealt{NIRCAM}) and its wavelength coverage (reaching up to $\sim5$\micron{}) {\color{referee} make NIRCam} ideal for the identification of candidate galaxies at redshifts above ten. 
To date,  several $z>10$ galaxy candidates have been reported
\citep{Adams2022a,Atek2022,Castellano2022,Donnan2022,Finkelstein2022,morishita22,Naidu2022,Yan2022} 
in the public extragalactic fields conducted with the NIRCam camera, including the Cosmic Evolution Early Release Science (CEERS) Survey \citep{Finkelstein2022CEERS}, the GLASS-JWST survey (\citealt{Treu2022a}), and the observations around the galaxy cluster SMACS J0723.3-7327 taken as part of the JWST-ERO. 
The unexpected abundance of high-redshift galaxies -- particularly at bright luminosities -- {\color{referee} could} be in tension with predictions from widely-adopted galaxy formation models \cite[e.g.,][]{Boylan-Kolchin2022,FPD2022,Finkelstein2022,Harikane2022,Labbe2022,MTT2022}. 

It is important to stress, however, that {\it none} of the $z>10$ candidates discovered by JWST have been spectroscopically confirmed to date {\color{referee} \citep{Fujimoto2022,Yoon2022}} and that the robustness of some of them have been called into question (e.g., \citealt{Naidu2022b,Zavala2022a}). 
Spectroscopic confirmation is thus necessary to measure the current tension between models and observations. 

The galaxy GHZ2/GLASS-z12 was first reported by \citet{Castellano2022} and \citet{Naidu2022} and is centered at RA = $+$00:13:59.76 DEC = $-$30:19:29.1. This source stands out as one of the best galaxy candidates at $z>10$ ever detected. Its photometric redshift ($z=11.960-12.423$ at a confidence limit of 68\%) has been confirmed by multiple independent teams \citep{Castellano2022,Donnan2022,Harikane2022,Naidu2022}, with negligible chances of being a lower-redshift contaminant due to its accurately measured colors and
sharp break in the NIRCam photometry. The depth of this feature, associated with the Lyman break, means that the redshift identification is still robust after the recent in-flight re-calibration of the JWST instruments \citep{Rigby2022}.

Here we present deep Atacama Large Millimetre/submillimetre Array (ALMA) spectroscopic and continuum observations towards this galaxy. 
This paper is organized as follows. In Section~\ref{sec:JWST} we briefly recap JWST observations for convenience of the reader. In Section~\ref{sec:2}, we discuss the ALMA observations, and we discuss the observational results {\color{referee} of the redshift identification using extensive statistical analysis} in Section~\ref{sec:3}. We discuss the implications of the redshift, line emission and statistical verification in Section~\ref{sec:4}, and provide future perspectives on the spectroscopic follow-up of JWST targets in Section~\ref{sec:5}. Finally, we summarize our results in Section~\ref{sec:6}.
Throughout this paper, we assume a flat $\Lambda$-CDM cosmology with $\Omega_\mathrm{m} = 0.3$, $\Omega_\mathrm{\Lambda} = 0.7$ and $h = 0.7$.

\section{Summary of JWST observations}
\label{sec:JWST}

The GLASS-JWST program represents the deepest extragalactic survey of the ERS campaign and consists of NIRISS \citep{Roberts2022} and NIRSpec spectroscopy observations centered on the cluster A2744 with parallel NIRCam imaging offset from the cluster center. The multi-band strategy of the NIRCam observations \citep{Merlin2022}, which include imaging in seven wide filters (F090W, F115W, F150W, F200W, F277W, F356W, and F444W), allows the identification of $z>10$ galaxy candidates via color-color diagrams and/or SED fitting techniques. The NIRCam  images used in this paper were reduced as described by \cite{Merlin2022}, who constructed a multi-band photometric catalog.  High-$z$ candidates were selected by \cite{Castellano2022} using a combination of color cuts and photometric redshifts designed to minimize contamination by lower redshift interlopers.

As mentioned above, GHZ2/GLASS-z12 was identified as a $z\sim12.5$ candidate by several teams using independent reductions of the GLASS data \citep{,Donnan2022,Harikane2022,Naidu2022}. \citet{Santini2022} presented the physical properties of this galaxy, which we update here using the most recent  photometric calibrations \citep{Rigby2022}. From our best-fit SED we constrain the following physical properties: a star formation rate of $\rm SFR=19^{+14}_{-10}\,\rm M_\odot\,yr^{-1}$, $M_\star=1.6^{+1.9}_{-0.3}\times10^{8}\,\rm M_\odot$, and absolute magnitude M$_{1500}=-21.0^{+0.2}_{-0.2}\,{\rm AB}$ \citep{Santini2022}.

\section{ALMA Observations and Data reduction}
\label{sec:2}
\begin{figure*}
	\includegraphics[width=\linewidth]{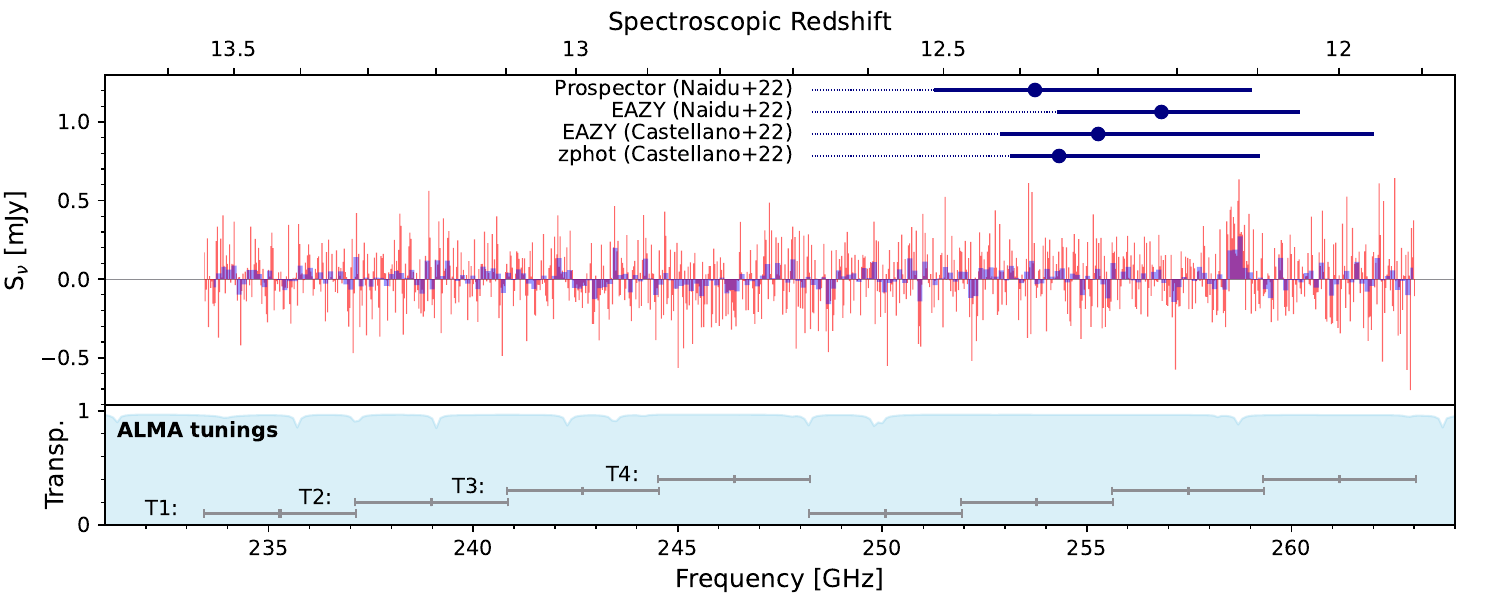}
\caption{{\it Top:} The full ALMA spectrum covers 233.42 to 263.04~GHz across four tunings of GHZ2/GLASS-z12. The {\it red} and {\it blue fill} show the spectrum at 35~km/s and 150~km/s bins, respectively. An emission feature is seen at 258.7~GHz 0\farcs5 north-east of the JWST position, extended across $\sim 0.4$~arcsec. Associating the line with \oiiil{}, the emission line confirms the spectroscopic redshift of GHZ2/GLASS-z12 to be $z = 12.117$. 
{\it Bottom:} The atmospheric transmission at 0.5~mm precipitable water vapour -- similar to the ALMA observing conditions (see Table~\ref{tab:alma_observations}) -- shows only minor absorption features ($< 10$~\%). The four tunings span the redshift range 11.9-13.5, covering 98\% of the confidence limits predicted from multiple photometric redshift methods \citep{Castellano2022,Naidu2022}}
    \label{fig:spectrum_oiii}
\end{figure*}

ALMA observations were carried out between 2022-08-03 and 2022-08-05 as part of the DDT project 2021.A.00020.S (Bakx \& Zavala), and are summarized in Table~\ref{tab:alma_observations}. The spectral setup consists of four adjacent tunings covering a total bandwidth of $\sim30\,\rm GHz$ from 233.4 to $263.0\,\rm GHz$ in the ALMA Band 6. This range covers the expected (redshifted) frequency of the [OIII]\,88\micron{}\ ($\nu_{\rm rest}=3393.0062\,\rm GHz$) from $z=11.9$ to $z=13.5$, where our target was expected to be (covering $\sim98$\% of the posterior distribution function of the photometric redshift).  Each of the tunings was observed for around $2.2\,$hrs on-source ($\sim4\,$hrs per tuning including the overheads).

{\color{referee} Based on the photometric redshift analysis by \citet{Castellano2022} and \citet{Naidu2022} conducted with {\sc eazy} and {\sc zphot}, we expect only a 2\% chance that the line be redshifted below or above our observing window. 
The initial results from the  
{\sc prospector} fit \citep{Naidu2022} suggest a slightly greater probability within $z\approx13.5-14.5$, although improved photometric estimates have now removed this redshift solution.
In addition, potential systematic errors in the photo-$z$, or selection effects altering the prior distribution could lead to underestimating these probabilities. Despite this, we believe the chances of the redshift being outside our window of observation are minor because the marginal detection in F150W and the clear photometric break tightly constrain the photometric redshift regardless of the prior and template choice. }

Data reduction was performed following the standard procedure and using the ALMA pipeline. Then, we use CASA for imaging the {\it uv}-visibilities using Briggs weighting with a robust parameter of 2.0 (to maximize the depth of the observations at the expense of slightly increasing the final synthesized beam size). 
This process results in a typical depth of $0.1\,$mJy\,beam$^{-1}$ in 35\,km\,s$^{-1}$ channels with a mean synthesized beam size of $\theta\approx0\farcs34\times0\farcs30$. In addition,  to have a
better sensitivity to extended emission beyond the $\sim 0.3$~arcsecond beam and to broad emission lines,  
we explore {\it uv}-tapering  at 0.3, 0.5, and 1.0~arcseconds and we create several cubes varying the velocity binning across the full frequency coverage, creating cubes with 15, 50, 100, 150, 300, and 400~km/s channels. Finally, we combine the four different tunings to create a single continuum image (at a representative frequency of $\nu_{\rm obs}=248\,$GHz) adopting Briggs weighting with a robust parameter of 2.0. The final continuum image has a root-mean square of $4.6\,\mu\rm Jy\, beam^{-1}$ and a beam size of $\approx0\farcs34\times0\farcs31$.

\section{
Line search and dust continuum emission}
\label{sec:3}
We look for the emission of \oiii{} at and surrounding the JWST position of GHZ2/GLASS-z12 using different velocity binnings (including velocity offsets) and taperings. We find an emission line offset from the source by a projected $\sim 0.5$~arcseconds and perform extensive statistical tests to verify its robustness. We further discuss the 
properties of this detection and its potential caveats, as well as, the 
lack of any dust or line emission at the source position.

\subsection{The [O\,{\sc iii}] emission line from GHZ2/GLASS-z12 at $z=12.117$} 
\label{sec:tentativelinesection}
We find a moderately-extended $5.8 \sigma$ feature at 
$0\farcs5$ north-east of the JWST source at $\sim258.7$~GHz, which we associate with \oiii\,88 \,$\mu$m  at $z = 12.1$.  At this redshift, the position offset (which is larger than the expected absolute astrometric accuracy of $<0\farcs1$) corresponds to a physical offset of $\sim 1.5$~kpc. The full 30\,GHz spectrum at this position is shown in Figure \ref{fig:spectrum_oiii}, while a zoomed-in version of the line profile can be seen in Figure~\ref{fig:tentativeline}. 
This line feature 
spatially extends across $0\farcs4$. Using the {\sc emcee} Monte Carlo fitting tool \citep{emcee2013}, we extract the line properties, which is centered at $258.68 \pm 0.03$~GHz and has a total velocity-integrated line intensity of $0.193 \pm 0.036$~Jy~km/s, with a line full-width at half-maximum of $400 \pm 70$~km/s. The spectroscopic redshift associated with this line detection is $z = 12.117\pm0.001$ and the line luminosity is $L_{\rm [OIII]}=9.0 \times{} 10^8$~L$_{\odot}$ (following \citealt{Solomon2005}).

\subsubsection{Observational tests to verify the emission line}
\begin{figure}
    \centering
    \vspace{-0.3cm}
    \includegraphics[width=\linewidth]{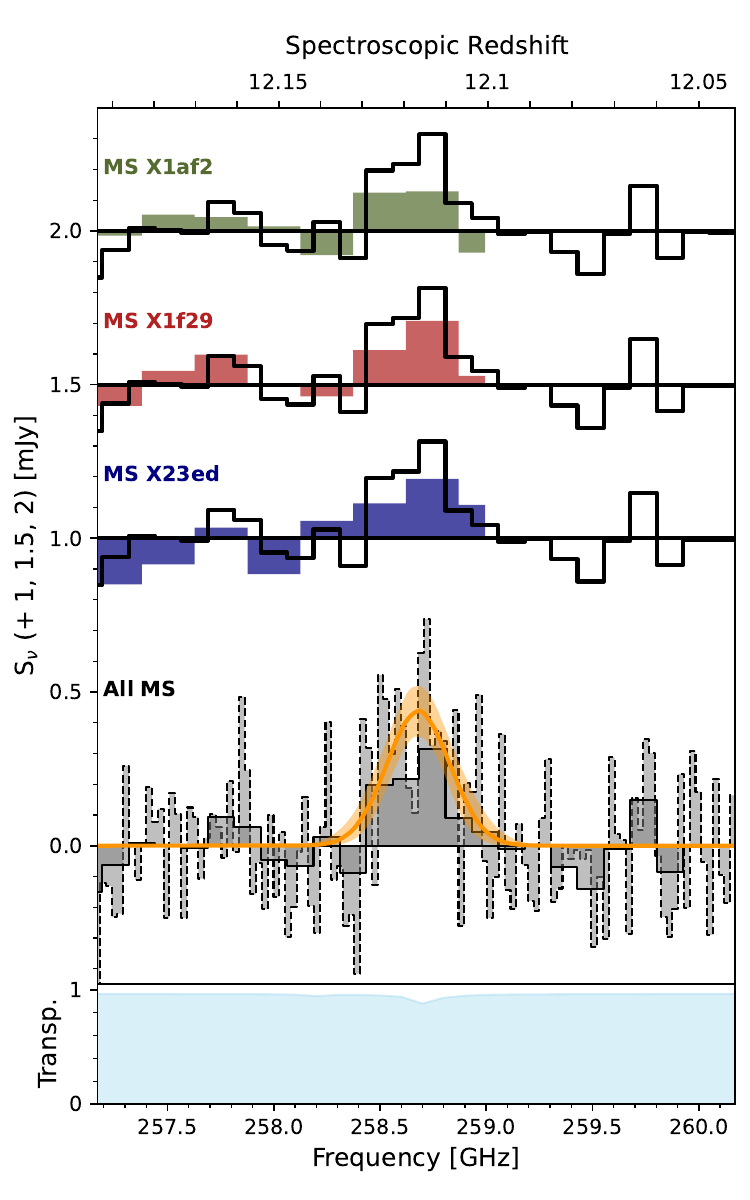}
    \caption{The emission line  at $z=12.117$ from GHZ2/GLASS-z12, seen in three separate different measurement sets (see Tuning 3 in Table~\ref{tab:alma_observations}) and in the combined image at 150~km/s ({\it solid lines}) and at 35~km/s ({\it dashed lines}).  The {\it orange line and fill} indicate the {\sc emcee}-based line fit and errors on the line fit. The emission is seen in the three different executions, although with a lower signal-to-noise ratio, as expected. The {\it bottom panel} indicates the atmospheric transparency, which shows a feature around the position of the line. Around the line, the atmospheric transmission is $\sim 95$~per cent, while it drops to $\sim 90$~per cent near the feature. The line fit has a significance of $5.5\sigma$, and $5.8 \sigma$ in the statistical analysis using a manually-adjusted aperture (see Figure~\ref{fig:gaussian}).}
    \label{fig:tentativeline}
\end{figure}

To assess the reliability of this detection, we first check the emission across  the three independent executions covering this frequency range (Tuning 3 from Table~\ref{tab:alma_observations}). Marginal emission is seen in the three different observations (Figure \ref{fig:tentativeline}), disfavoring a
false-positive associated with a single noise spike. Instead, the fact that the emission feature is seen across all three tunings further improves the probability of this being a true line detection. 
We note that this emission lies in the middle of an atmospheric absorption feature, which could boost the noise at the frequency of the observed line. Nevertheless, the atmospheric transmission is still very high (close to $\sim 90$~per cent as shown in the bottom panel of Figure \ref{fig:tentativeline}) and its effect is thus expected to be small.

\subsubsection{Statistical tests to verify the emission line}
\label{sec:statisticalTests}
\begin{figure}
    \centering
    \vspace{-0.3cm}
    \includegraphics[width=\linewidth]{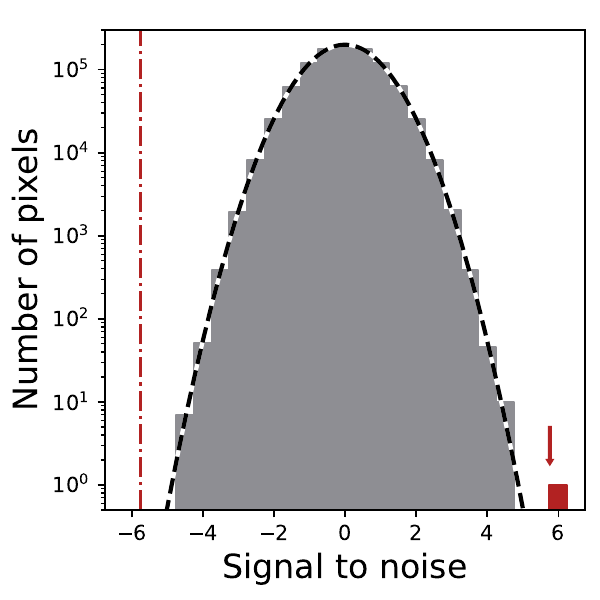}
    \caption{The emission line ({\it red bin and arrow}) stands out by 5.8~$\sigma$ from the three-dimensional data cube compared to one million normalized flux extractions from off-line positions in both the positive and negative significance distribution ({\it dash-dotted red line}). Each frequency slice in the data cube is normalized to the standard-deviation prior to extracting a manually-optimized aperture across the emission line (shown as a dotted box in the left-most panel of Figure~\ref{fig:momentsmap}). The resulting relative signal-to-noise in the emission line is then normalized by the Gaussian fit of the one million off-line extractions. This approach further allows us different apertures to test the robustness of the line, with the results shown in Apppendix Figure~\ref{fig:appendix2Dintegrallimits}. }
    \label{fig:gaussian}
\end{figure}

We perform an in-depth statistical analysis to estimate the veracity of the line emission through a comprehensive Monte Carlo simulation. 
We use a 0.3 arcsecond tapered data cube with a velocity sampling at 150~km/s.
We normalize the entire data cube to the per-frequency standard deviation to account for the inhomogeneous noise-profile of the emission due to observational and atmospheric effects. We then manually define a square aperture in both x-, y- and frequency pixels. 
Here we mask out a single bright emission line in the north-west of the cube associated with a bright foreground galaxy, and proceed to take one million samples across the data cube at off-line positions. 
We then fit the relative signal-to-noise distribution of all the one million measures    with a Gaussian profile,  to have an estimate of the normalized  noise distribution across the whole data cube, taking into account the aperture size effects. This would account for any coherent noise in the system missed in either direct line fitting or 2D fitting. As shown in Figure~\ref{fig:gaussian}, the normalized signal-to-noise of our signal is $5.8 \sigma$, with no single other aperture matching the emission at both positive and negative signal-to-noise, confirming the robustness of the line. 

In the Appendix~\ref{sec:appendixVariableVelocity}, we expand upon this analysis in order to investigate the wide line-width of the line. There, we try the line fitting for different frequency bounds on the aperture. Appendix Figure~\ref{fig:appendix2Dintegrallimits} shows the effect of changing the integration velocities from -450 to +750 at 150~km/s intervals for a total of 36 different integration configurations. Even with a 150~km/s line velocity, we find a $> 5 \sigma$ detection and a total of six such configurations resulting in a line significance in excess of $5 \sigma$. 

\subsubsection{On the line-width, size, and spatial offset}
\begin{figure*}
    \centering
    \includegraphics[width=\linewidth]{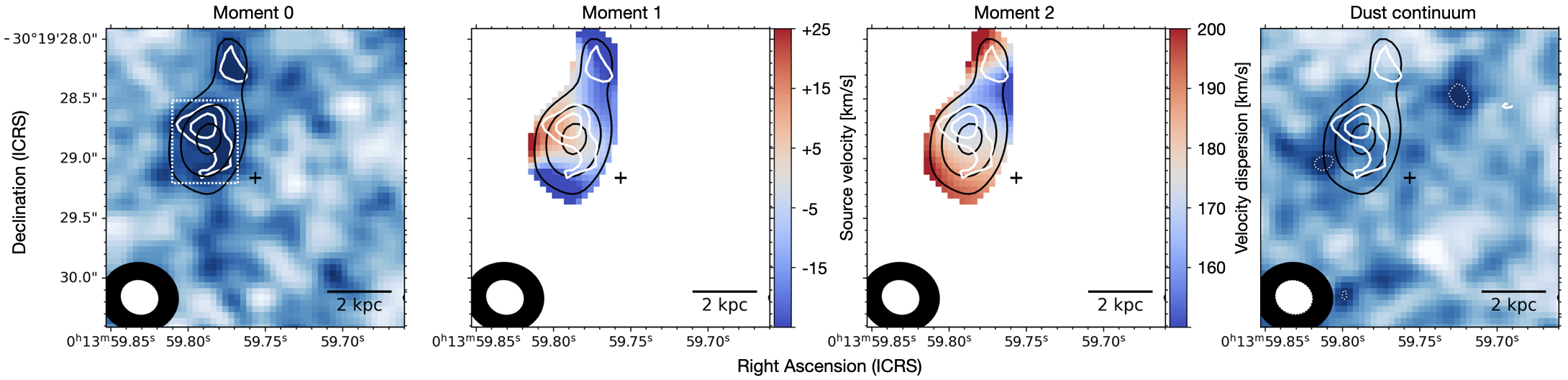}
    \caption{The line associated with GHZ2/GLASS-z12, seen untapered with robust = 2 ({\it white contours}) and tapered at 0.5~arcseconds ({\it black contours}), drawn at $\pm3, 4, 5\sigma$ levels. The {\it central black plus} indicates the JWST position. 
    {\it Left:} The line emission (moment-0 map) is offset from the JWST-source and appears extended. The dashed box indicates the region used for the analysis in Figure~\ref{fig:gaussian} discussed in Subsection~\ref{sec:statisticalTests}. 
    {\it Middle left:} The velocity gradient (moment-1 map) of the line shows little gradient in the velocity profile of the line.
    {\it Middle Right:} The velocity dispersion (moment-2 map) of the line shows an average velocity dispersion of 180~km/s across the source.
    {\it Right:} No dust emission is seen at the JWST position, nor at the position of the spectral line. 
    }
    \label{fig:momentsmap}
\end{figure*}
The emission line is significant, however here we note several caveats: {\bf (1)} the line is spatially-offset from the JWST detection. {\bf (2)} The large velocity width is in excess of what is seen for systems with stellar masses of $\sim10^8$~M$_{\odot}$. For comparison, \cite{Inoue2016,Laporte2017,Laporte2021,Hashimoto2018,Tamura2019} report values between 50 and 320~km/s.
And finally, {\bf (3)} the emission appears spatially more extended than the size inferred from the JWST images of GHZ2/GLASS-z12 \citep{Yang2022}.

{
While these line properties are surprising for a $z \sim 12$ galaxy, we discuss some possible explanations. 
First, we note that spatial offsets between emission lines \oiii{}, \cii{}, Ly$\alpha$ and the UV or dust continuum have been reported both in observations \citep[e.g.,][]{carniani:2017oiii} and simulations \citep[e.g.,][]{Pallottini2019,Katz2019,Arata2020} at $z = 7 - 8$. These offsets are typically understood to be due to chemically-evolved components with high dust-obscuration or by outflows of chemically-enriched gas. 
Indeed, an outflow-scenario would be able to explain not only the large spatial offset but also the large line width and spatially-extended emission. Similarly, the observed line properties could be the result of a galaxy interaction.  In this case, it would require the presence of a heavily-obscured component to explain the non-detection in the NIRCam filters and a weak dust emission contrast against the CMB to explain the non-detection of dust continuum (e.g., \citealt{daCunha2013,Zhang2016}, see Section~\ref{sec:dust}). To further explore these possibilities, we show the moment zero, one and two maps of the emission line  in
Figure~\ref{fig:momentsmap}, as well as the map of undetected dust emission. The emission line is offset by 1.5~kpc, and has a clumpy structure extending to the north. A modest velocity gradient ($\sim 15$~km/s) appears in the direction away from the JWST source. Meanwhile, the velocity dispersion of the emission line varies little across the emission line region. There are no indications of rotation, while the velocity gradient could be caused by a decelerating outflow. 
Although it is certainly possible that early phases of galaxy evolution are dynamically complex \cite[e.g.,][]{Arata2019,Ziparo2022}, making these scenarios conceivable, further observations of the peculiar nature of this emission line are needed to discern between these various interpretations.

Finally, it is worth noting 
the relatively large uncertainties in the measured line velocity and spatial extension. Therefore, it is also possible that the true line velocity could be lower given the relatively-large errors in the velocity width ($400 \pm 70$~km/s) and line significance even at smaller integration velocities (Appendix Figure~\ref{fig:appendix2Dintegrallimits}). The same is true for the extended emission which is only marginally larger than the beamsize (see Figure \ref{fig:momentsmap}).
}

\subsection{The lack of \oiii{} emission at the position of GHZ2/GLASS-z12} 
\label{sec:upperlimit}
\begin{figure}
    \centering
    \vspace{0.3cm}
    \includegraphics[width=\linewidth]{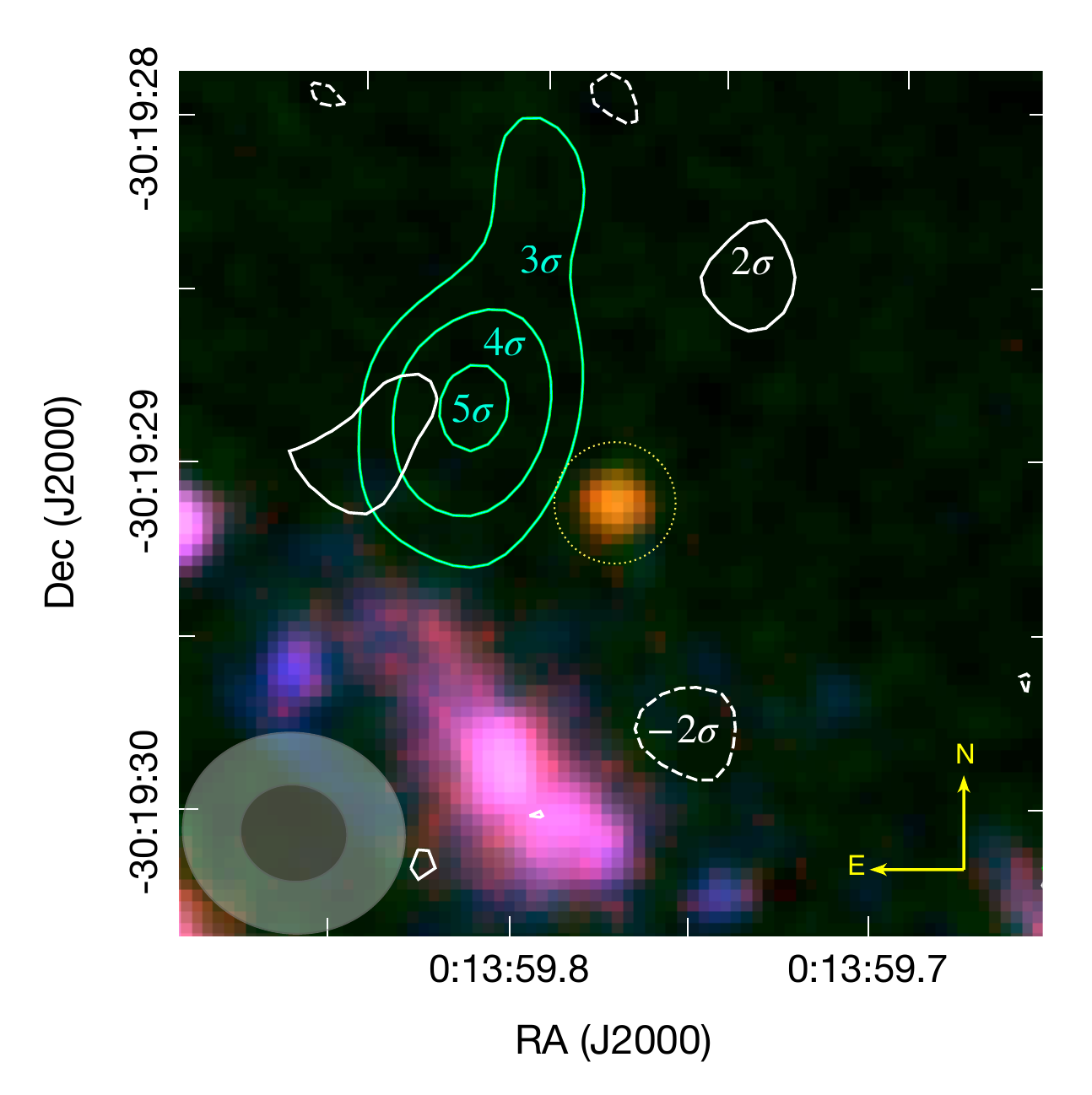}
    \caption{A $2\farcs5\times2\farcs5$ JWST/NIRCam composite image of GHZ2/GLASS-z12 is shown in the background (F150W in {\it blue}, F277W in {\it green}, and F444W in {\it red}) along with the dust-continuum signal-to-noise ratio as {\it white contours}. Since there is no dust continuum emission above $\pm3\sigma$, only $\pm2\sigma$ contours are shown.
    To illustrate the offset and the significance of the tentative emission line at $z=12.117$, we also plot $\pm3$, 4, and $5\sigma$ levels of the moment-0 map (with 0.5~arcsec tapering) across 258.5 to 259.0\,GHz as {\it green contours}.   The beam sizes for the (untapered) continuum map and the tapered moment-0 map are represented by the dark and light ellipses on the bottom left. The $0\farcs35$ aperture used to extract the upper limit at the JWST position is also illustrated with a yellow dotted circle.
    }
    \label{fig:stamps}
\end{figure}

No obvious emission line is seen at the JWST position of GHZ2/GLASS-z12, as shown in Appendix Figure~\ref{fig:spectrum_plus_tentative}. The spectrum of GHZ2/GLASS-z12 extracted from an aperture centered on the JWST-position with a circular size of $0\farcs35$ selected to match the average synthesized beam size. We show this aperture relative to the background JWST image in Figure~\ref{fig:stamps}. Similarly, no emission is seen in any of the resampled spectra with different velocity binnings (including velocity offsets) and taperings.

In order to evaluate the intrinsic properties of the UV-bright component of GHZ2/GLASS-z12, we also estimate an \oiii{} line luminosity upper limit at the the exact JWST position. 
We use the standard-deviation of the map at each frequency to evaluate the \oiii{} luminosity upper limit. We find no redshift dependency, although the atmospheric windows and instrumental sensitivity slightly vary across the spectral windows. The average $5\sigma$ line luminosity upper limit across the entire window is estimated to be  $1.7 \times{} 10^8$~L$_{\odot}$ assuming a line velocity of 100~km/s and no spatially-extended emission. Assuming a wider line-width of 200~km/s would increase the derived upper upper limit by $\sim40\%$.


\subsection{Search for dust continuum emission} \label{sec:dust}
No dust emission is seen in the collapsed (multi-frequency synthesis) continuum image, down to $13.8$~$\mu$Jy at $3 \sigma$. The lack of dust emission provides further credence to the high-redshift solution at $z = 12.117$. Assuming a typical dust thermal emission SED (e.g., \citealt{Casey2012}), we derive an upper limit on the dust-obscured star formation of $< 2 - 5$~M$_{\odot}\,\rm yr^{-1}$ at $3 \sigma$ for low-redshift interlopers ($z < 6$), depending on the galaxy model. Hence, these observations rule out the possibility of a low-redshift interloper associated with a dusty star-forming galaxy where the observed break in the NIRCam photometry would be rather associated to the Balmer break combined with high dust attenuation (e.g., \citealt{Zavala2022a}).

The dust non-detection is fully consistent with the blue colors and multiple JWST detections redwards of the strong Lyman break, which also rule out a $z\sim4$ quiescent galaxy. Furthermore, the compact size of  $0.047\pm0.006\,$kpc (corresponding to $0.17\pm0.02$\,kpc at $z\sim12$; \citealt{Yang2022}) is much more compatible with a high-redshift source than with a one at much lower redshift. The contamination from a dwarf star has also been ruled out since dwarf SED templates do not provide a good fit to the NIRCam data points, moreover the source is clearly resolved. Again, this is consistent with a $z = 12.117$ identification for GHZ2/GLASS-z12.

\section{Discussion}
\label{sec:4}
\subsection{Metallicity and the \oiii{}-SFR relation}
\label{sec:metallicity}
Figure~\ref{fig:oiii_sfr} shows the \oiii{} emission line and the upper-limit of \oiii{} emission at the position of GHZ2/GLASS-z12 against the star-formation estimate from JWST observations. The line emission and the on-source upper-limit from Subsection~\ref{sec:tentativelinesection} and \ref{sec:upperlimit} are compared to local starbursting galaxies \citep{delooze14}, metal-poor galaxies \citep{Cormier2019,Harikane2019} and a reference sample of $z > 6$ Lyman-break selected galaxies from \cite{Harikane2019}.
Below we discuss the interpretation of such measures and the derived constraints on the gas-phase metallicity.


\subsubsection{Metallicity estimate of the line-emitting region}
\begin{figure}
    \centering
    \includegraphics[width=\linewidth]{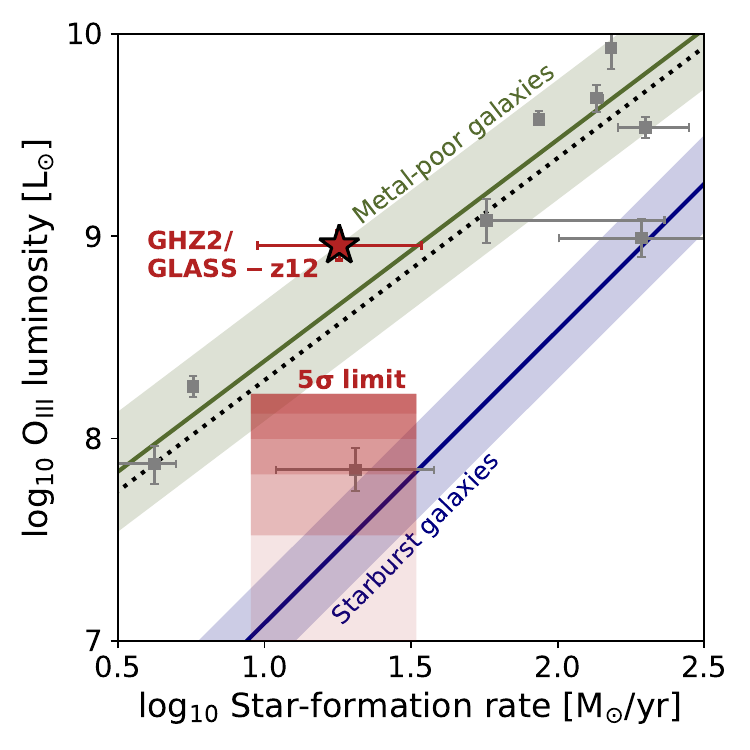}
    \caption{Star-formation rate of distant galaxies as a function of \oiiil{} emission. The line emission lies at the top end of the scaling relations from metal-poor galaxies (\citealt{Cormier2019, Harikane2019}), and the range seen for $z \approx 6-9$ galaxies (gray squares; \citealt{Harikane2019} and references therein). The {\it red fill} indicates our 5 through $1 \sigma$ upper limit on the Oxygen luminosity at the position of the JWST emission. The upper limit coincides with observed distant galaxies, as well as the scaling relation for starburst galaxies (\citealt{delooze14}). }
    \label{fig:oiii_sfr}
\end{figure}

{
As shown in Figure \ref{fig:oiii_sfr}.
the line detection lies slightly above the scaling relation for metal-poor galaxies when adopting the JWST-based SFR of $19^{+14}_{-10}\,\rm M_\odot\,yr^{-1}$ (\citealt{Santini2022}; although still consistent within the error bars). This could suggest an enhancement of \oiii{} emission in this system.

If we use instead the 3$\sigma$ limit of SFR~$< 11$ M$_{\odot}$~yr$^{-1}$ based on non-detection of dust emission at the position of the line emission and, following Equation~2 of \cite{Jones2020}, the emission line corresponds to a 3$\sigma$ {\it lower} limit on the metallicity of $12 + \log{O/H} > 8.8$, i.e., a super-solar oxygen abundance
(adopting  electron temperature $T_e = 1.5 \times 10^4$~K, gas density $n_e = 250$~cm$^{-3}$, and an ionization correction factor of 0.17 dex from O$^{++}$ to total Oxygen abundance, with solar metallicity $12 + \log{O/H}_{\odot} =8.69$; \citealt{Asplund2009}).
However, the uncertainty arising from unknown physical conditions is of order 0.4 dex, which could significantly reduce the lower limit on the metallicity. Assuming extreme nebular densities and temperatures ($n_e = 1$~cm$^{-3}$, $T_e = 2.5 \times 10^4$~K), the associated abundance would be half the solar value (i.e. $12 + \log{O/H}_{\odot}>8.4$).

A high metallicity is surprising given the lack of any stellar emission at the position of the line-emitting region, particularly at this high redshift. And while some recent studies have suggested
an early onset of star formation and a rapid evolution in $z>10$ galaxies \cite[][]{Boylan-Kolchin2022,FPD2022,Finkelstein2022,Harikane2022,Labbe2022,MTT2022,PG2022}, it may suggest that this line does not arise from star-forming \hii\ regions (e.g., with ionized outflows as an alternative scenario instead; e.g., \citealt{Fiore2022,Ziparo2022}). The high metallicity estimate from the wide, offset emission line is also affected by the star-formation rate estimates (based on a $\sim 50$~K dust temperature) and the correct line velocity. These affect the metallicity linearly, with both an increase in star-formation and a decrease in line width decreasing the estimated metallicity. Similarly, the assumed electron temperatures, gas densities and O$^{++}$-to-Oxygen abundances might vary, even relative to the $z = 6-9$ Universe.
}

\subsubsection{Metallicity estimate at the JWST position}
In contrast to the high metallicity associated with the emission line, the 5$\sigma$ line flux limit implies an oxygen abundance $12 + \log{O/H} < 7.6$. Same as above, the uncertainty arising from unknown physical conditions is of order 0.4 dex \citep{Jones2020}. This limit corresponds to $< 0.1$ times the solar value, and is comparable to the typical metallicities inferred for luminous \oiii\ emitters at $z\sim8$ \citep{Jones2020}. 

Our metallicity limit implies that the JWST-visible component of GHZ2/GLASS-z12 is likely to be in an early stage of chemical enrichment. 
From a simple closed-box chemical evolution model, assuming oxygen yields $y_O = 0.007-0.039$ from low-metallicity stars \citep{Vincenzo2016}, the metallicity of GHZ2/GLASS-z12 suggests only $< 2$--14\% of its gas has been processed into stars (i.e. $>90$\% gas fraction; the constraint becomes $> 82$\% for the case of a 400~km/s line width). However, effects of gaseous inflows and outflows can permit smaller gas fractions; the low metallicity may thus indicate accretion and outflow rates which are comparable or larger than the SFR. 
In any case the non-detection of \oiii{} at the JWST position suggests that the metal abundance of GHZ2/GLASS-z12 might not yet be as high as that seen in $z = 6 - 9.2$ Lyman Break Galaxies \citep{Harikane2019,Jones2020}. This is expected given that only $\sim 400$ Myr elapsed from the Big Bang to the time of observation, leaving little time to form heavy elements \citep{Maiolino2019,Ucci2021}. Our low metallicity limit further corroborates the young age based on SED fitting \citep{Santini2022}. 

\subsubsection{Observed metallicity gradient across GHZ2/GLASS-z12}
The large variation in star-formation and oxygen-emission properties of the line-emitting and JWST-observed regions of GHZ2/GLASS-z12 suggest a metallicity gradient exists across the source (if the \oiii{} emission arises from star-forming \hii{} regions), even considering the uncertainties in the metallicity estimates since we need to assume many galaxy properties. Previous observations at lower redshift suggest pre-existing stellar populations formed at redshifts $z > 10$ enrich galaxy systems \citep{Hashimoto2018,Hoag2018,Tamura2019,RobertsBorsani2020,PG2022}, as well as episodic star-formation \citep{Arata2019,Katz2019,Pallottini2019} distributing the chemicals efficiently \citep{Sun2022,Ziparo2022}. The detection of galaxies by strong Lyman-breaks further selects towards young stellar populations, which might not spatially coincide with these older, enriched populations.

\subsection{Lack of dust in the cosmic dawn?}
The lack of a dust detection (down to a $3 \sigma$ limit of 13.8~$\mu$Jy; see Figure~\ref{fig:stamps}) suggests an upper limit of $1.5\times{}10^6$~$M_{\odot}$ of inter-stellar dust, a far-infrared luminosity less than $6.5 \times{} 10^{10}$ L$_{\odot}$, and a dust-obscured star-formation rate of 11~M$_\odot$\,yr$^{-1}$. This explains the blue UV slope ($\beta_{UV} \approx -2.4$), suggesting little dust obscuration of the young ($\sim 70$~Myr; \citealt{Naidu2022}) stellar population. This assumes a dust temperature of 50~K, although average temperatures could rise to 75~K or beyond based on the observed dust temperature evolution with cosmic distance reported in e.g., \cite{Bouwens2020,Bakx2021}. 
The dearth of dust is in line with dust production models, which typically require several tens of Myr before the supernovas of the heaviest stars produce the metals necessary for dust. Wolf-Rayet stars are an alternative dust production pathway, where the orbital dynamics of two binary stars creates a region where stellar winds are able to produce dust \citep{Lau2022}. We can place a relatively weak constraint on the dust production from these types of systems down to $< 1.5 \times{} 10^{-3}$ M$_{\odot}$/star, in line with models by \cite{Lau2021}.

\begin{figure}
    \centering
    \includegraphics[width=\linewidth]{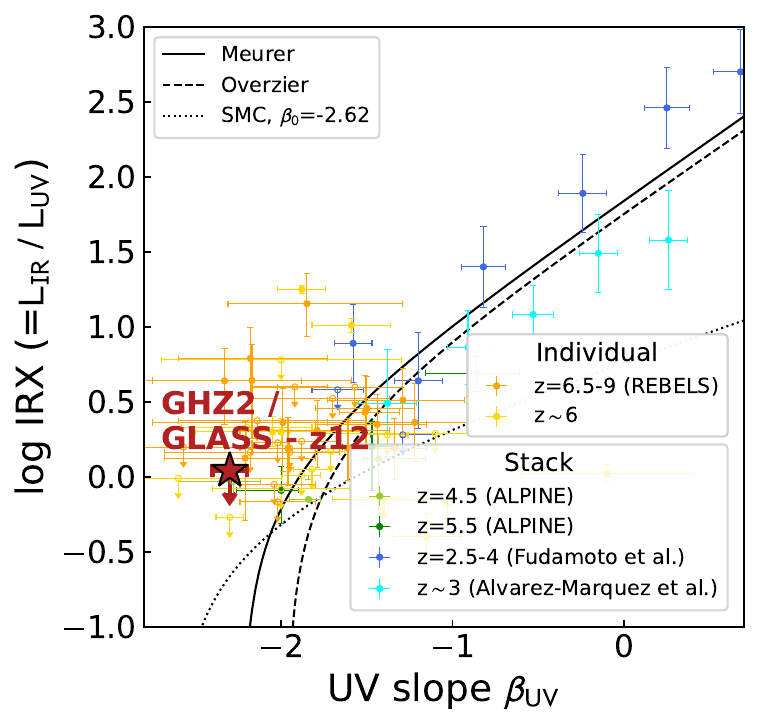}
    \caption{The IRX-$\beta_{\rm UV}$ relation of GHZ2/GLASS-z12 suggests that our $3 \sigma$ dust upper-limit is deep, probing some of the most dust-unobscured systems ever detected. The source is compared against reference samples from ALPINE \citep{Fudamoto2020}, REBELS \citep{Bouwens2021Rebels,Inami2022}, and $z \approx 3$ galaxies \citep{AlvarezMarquez2019}, as well as theoretical models by \citet{Meurer1999} and \citet{Overzier2011}. }
    \label{fig:IRX_beta}
\end{figure}

Figure~\ref{fig:IRX_beta} shows the comparison of the dust-obscured emission ($\rm IRX = L_{\rm IR}/L_{\rm UV}$) against the UV slope, and finds that GHZ2/GLASS-z12 is at the low end of dust-obscured star-formation. The $\log_{10} \rm IRX$ can move upwards by 0.5 if the dust temperature is 75~K instead of 50~K, removing the source from the extremely low IRX region. Regardless, this galaxy stands in contrast to the relatively high dust-obscuration factors found at $ z\sim8$ (e.g., \citealt{Inami2022}), implying a very low dust content in the early Universe and a negligible dust attenuation at $z\sim12$. This is consistent with the recent calculations by \citet{MTT2022} and \citet{FPD2022}, who concluded that a negligible dust attenuation is necessary to explain the number of bright JWST candidates  reported at $z\approx11-14$.

\section{Future prospects for spectroscopy of $z>10$ galaxies}
\label{sec:5}
{\color{referee}
We report a line and, associating
it with the \oiiil{} line, infer a spectroscopic redshift of $z = 12.117 \pm 0.001$. In this section we present some lessons for distant redshift searches that may help guide follow-up of GHZ2/GLASS-z12 and spectroscopic confirmation of other high redshift sources that are being found by JWST. }

As far as ALMA is concerned, with Carbon requiring nearly half a billion years to build up \citep{Maiolino2019}, the typical Oxygen timescale (at 50~Myr) makes it generally the best spectroscopic redshift indicator. Indeed, as thoroughly explored in \cite{Bouwens2021Rebels}, and initially-indicated by \cite{Inoue2016}, \oiii{} is likely the brightest line in the distant Universe. {\color{referee}
The relatively narrow bandwidth of the ALMA receivers prevented us to cover the full photometric redshift probability distribution with a single tuning, necessitating a compromise between additional redshift coverage against at the cost of substantial overhead. 
The development of wider bandwidth receivers (\citealt{ALMAROADMAP,Carpenter2022}) would significantly speed up the process of building large samples of spectroscopically-confirmed galaxies at these early epochs, and the characterization of their metallicity and dust content, which remain a major and compelling scientific goal for ALMA. 
}

In the near infrared, JWST-NIRSpec should be able to provide conclusive redshift identification for large samples of galaxy candidates at {\color{referee} $z > 10$ redshifts identified by NIRCam (see e.g., \citealt{RB2022})}. For targets as bright as GHZ2/GLASS-z12, just a few hours of integration with the prism would be sufficient to detect the continuum, and thus secure a redshift identification via identification of the Lyman break at high spectroscopic resolution. If emission lines are present, the same short prism observations would detect common emission lines such as \ion{N}{V}$\lambda 1242$, \ion{C}{iV}$\lambda 1548$, \ion{He}{II}$\lambda 1640$, \ion{O}{III}]$\lambda 1660$, \ion{C}{III}]$\lambda 1909$ -- and [\ion{O}{II}]$\lambda\lambda3726,3729$ below $z\sim13$ -- for equivalent width as low as 5\AA.  The detection of these lines would nicely complement detection or upper limits on [\ion{O}{III}] from ALMA, in terms of metalliticy measurements \citep[see discussion by][at lower redshift]{Jones2020}. Even for candidates not as photometrically secure as GHZ2/GLASS-z12, with colors and photo-$z$ allowing for lower-redshift solutions, JWST-NIRSpec should easily distinguish the Lyman Break from the most likely contaminants, which are galaxies with the Balmer break at the corresponding wavelength and blue rest-frame optical colors, owing to the abundant and strong  lines around the Balmer break. 

At wavelengths between NIRSpec and ALMA, JWST-MIRI should provide a third important window into early galaxy formation, by allowing the detection of strong optical emission lines such as H$\beta$ H$\alpha$ and [\ion{O}{III}]$\lambda\lambda4959,5007$, if they are present and strong.

We conclude that ALMA and JWST are highly synergistic and together they should revolutionize our understanding of early galaxy formation and evolution.

\section{Summary}
\label{sec:6}
We reported on the ALMA band 6 redshift search for the spectroscopic redshift of GHZ2/GLASS-z12 through the \oiii{} emission line, covering 30\,GHz contiguously. 
{\color{referee} Our deep observations ($1\sigma=0.1\,$mJy\,beam$^{-1}$ in 35\,km\,s$^{-1}$ channels),
revealed a $5.8 \sigma$ line at 258.7~GHz and, associating
it with the \oiiil{} line, infer a spectroscopic redshift of $z = 12.117 \pm 0.001$. 

The projected offset nature of the line (0\farcs{}5 or 1.5~kpc) could be caused by an outflow or pre-existing but JWST-dark stellar components. Assuming star-forming \hii\ regions as the origin of the \oiii{} emission  requires
a high metallicity in the line-emitting region of $12 + \log{O/H} > 8.4$. At the JWST position, the \oiii{} luminosity upper-limit from our observations suggest a metal-poor system ($12 + \log{O/H} < 7.83$) in the distant Universe, with a 
lower line luminosity compared to $z\approx6-9$ galaxies. The lack of dust emission, even with our deep observations, contrasts with lower redshift galaxies, implying a very low dust content and a negligible dust-obscuration at this early epoch, potentially due to the short cosmic time.
}

We have also discussed potential strategies for deriving spectroscopic redshifts of $z\gtrsim11$ candidates, the necessity of improving current instruments' capabilities, and the importance of combining multi-wavelength observations to constrain the physical properties of the earliest galaxies in the Universe.

\section*{Acknowledgements}

This paper makes use of the following ALMA data: ADS/JAO.ALMA\#2021.A.00020.S. ALMA is a partnership of ESO (representing its member states), NSF (USA) and NINS (Japan), together with NRC (Canada), MOST and ASIAA (Taiwan), and KASI (Republic of Korea), in cooperation with the Republic of Chile. The Joint ALMA Observatory is operated by ESO, AUI/NRAO and NAOJ.
This work is partly based on observations made with the NASA/ESA/CSA James Webb Space Telescope. The data were obtained from the Mikulski Archive for Space Telescopes at the Space Telescope Science Institute, which is operated by the Association of Universities for Research in Astronomy, Inc., under NASA contract NAS 5-03127 for JWST. These observations are associated with program JWST-ERS-1324. We acknowledge financial support from NASA through grant JWST-ERS-1324. TB and YT acknowledge funding from NAOJ ALMA Scientific Research Grant Numbers 2018-09B and JSPS KAKENHI No. 17H06130, 22H04939. EV acknowledges financial support through grants PRIN-MIUR 2017WSCC32, 2020SKSTHZ and the
INAF GO Grant 2022 “The revolution is around the corner: JWST will probe globular cluster precursors 
and Population III stellar clusters at cosmic dawn” (PI Vanzella). We thank Stefano Carniani and Stefano Berta for their kind and useful discussions. {\color{referee} Finally, we would like to thank the anonymous referee for their insightful comments and suggested additions.}

\section*{Data Availability}
The data are publicly available through the ALMA science archive and the MAST portal managed by Space Telescope Science Institute. Other calibrated products used in this article will be shared upon request.


\bibliographystyle{mnras}
\bibliography{example} 



\appendix
\section{ALMA observation table}\label{secc:appendix}
In this appendix we summarise the ALMA observations, given in Table A1.

\begin{table*}
    \centering
    \caption{Parameters of the ALMA observations}
    \label{tab:alma_observations}
    \begin{tabular}{cccccc}
\hline
UT start time          & Baseline length & N$_{\rm ant}$ &  Frequency & T$_{\rm int}$ & PWV \\ 
$[$YYYY-MM-DD hh:mm:ss$]$  & [m] 			 & 			 	 &  [GHz]     & [min] &  [mm] \\ \hline
\multicolumn{6}{c}{\textbf{Tuning 1}} \\
2022-08-03 06:33:45    & 15 -- 1301 	 & 43  			 &  233.42--237.14 \& 248.22--251.94		  & 44.30 & 0.82\\
2022-08-03 07:48:41    & 15 -- 1301 	 & 43 			 &  233.42--237.14 \& 248.22--251.94		  & 44.37 & 0.94\\
2022-08-03 09:03:07    & 15 -- 1301		 & 43			 &  233.42--237.14 \& 248.22--251.94		  & 44.38 & 0.97\\
\multicolumn{6}{c}{\textbf{Tuning 2}} \\
2022-08-03 10:42:26	   & 15 -- 1301	     & 43  			 &  237.12--240.84 \& 251.92--255.64		  & 43.88 & 1.03\\
2022-08-03 12:03:52	   & 15 -- 1301	     & 43			 &  237.12--240.84 \& 251.92--255.64		  & 43.90 & 1.15\\
2022-08-04 06:50:26	   & 15 -- 1301	     & 44			 &  237.12--240.84 \& 251.92--255.64		  & 43.83 & 0.57\\
\multicolumn{6}{c}{\textbf{Tuning 3*}} \\
2022-08-04 08:14:20	   & 15 -- 1301	     & 44  			 & 	240.82--244.54 \& 255.62--259.34		  & 44.87 & 0.56\\
2022-08-04 09:32:36	   & 15 -- 1301	     & 44			 & 	240.82--244.54 \& 255.62--259.34		  & 44.85 & 0.57\\
2022-08-04 10:48:43	   & 15 -- 1301	     & 44			 & 	240.82--244.54 \& 255.62--259.34		  & 44.87 & 0.62\\
\multicolumn{6}{c}{\textbf{Tuning 4}} \\
2022-08-05 06:57:24	   & 15 -- 1301	     & 46  			 & 	244.52--248.24 \& 259.32--263.04		  & 44.05 & 0.48\\
2022-08-05 08:01:01	   & 15 -- 1301	     & 46 			 & 	244.52--248.24 \& 259.32--263.04		  & 44.02 & 0.45 \\ \hline
\multicolumn{6}{l}{{\bf *} Containing the \oiiil{} emission line at 258.7~GHz}\\
    \end{tabular}
\end{table*}

\section{Variable-frequency extraction of the emission line}
\label{sec:appendixVariableVelocity}
\begin{figure}
    \centering
    \includegraphics[width=\linewidth]{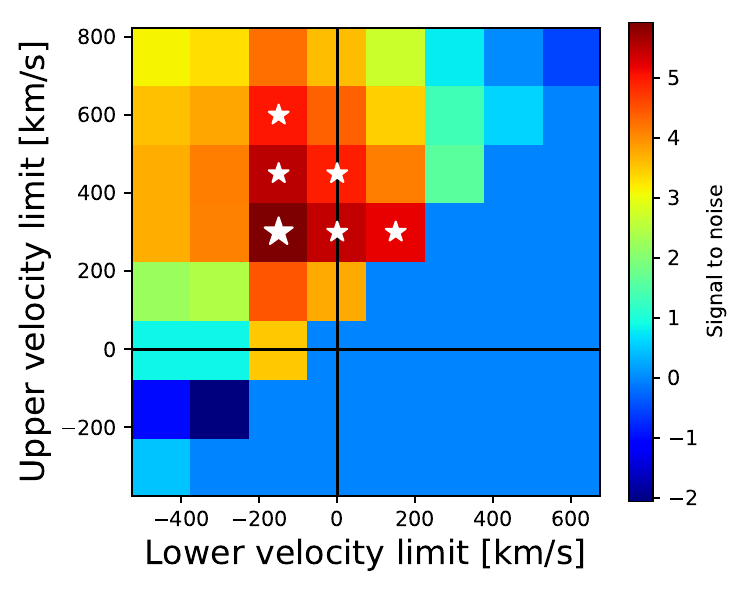}
    \caption{The line significance for different velocity integrals from the lower velocity limit ({\it x-axis}) to the upper velocity limit ({\it y-axis}). The line significance is indicated in a colour-scale ranging from $-2$ to $> 5 \sigma$. {\it Stars} indicate velocity integration bounds resulting in a significance in excess of $> 5 \sigma$, with the {\it larger star} indicating the maximum significance at $5.8 \sigma$.}
    \label{fig:appendix2Dintegrallimits}
\end{figure}

We apply the method of Subsection~\ref{sec:statisticalTests} assuming different velocity integration boundaries to test the veracity of the line, as shown in Figure~\ref{fig:appendix2Dintegrallimits}. The colour scale indicates the significance of the emission line, ranging from $-2$ to $> 5 \sigma$, when integrating from the lower velocity limit (x-axis) to the upper velocity limit (y-axis). There exist single velocity bins that are in excess of $5 \sigma$, i.e., from 150 to 300~km/s integration. The aperture was manually optimized for the -150 to 300~km/s, the highest significance bin, and indubitably the significance of the other bins could be improved with further manual optimization.

\section{Line spectrum at the JWST position}
We present the line spectrum at the JWST position, extracted from a 0\farcs{}35 aperture at the JWST position. No emission line above $4 \sigma$ is visible in this spectrum.
\begin{figure*}
	\includegraphics[width=\linewidth]{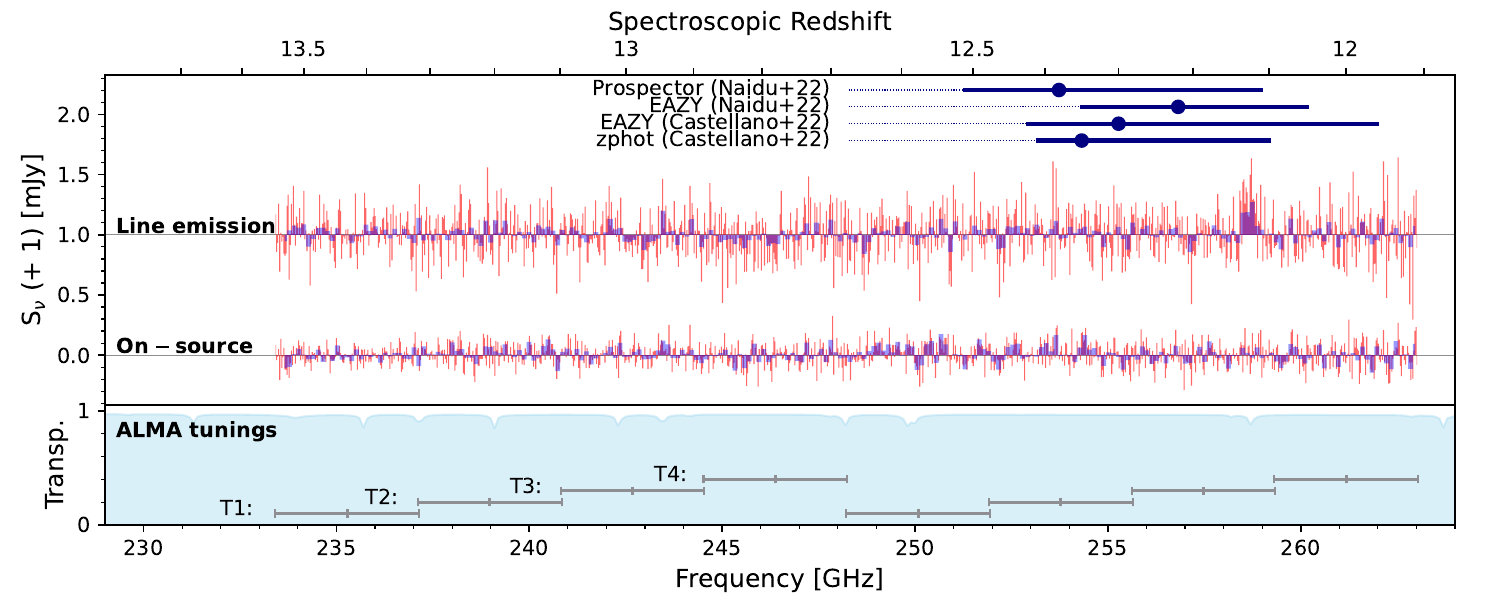}
\caption{Similar to Figure~\ref{fig:spectrum_oiii} {\it Top:} The full ALMA spectrum covers 233.42 to 263.04~GHz across four tunings of GHZ2/GLASS-z12. The {\it red} and {\it blue fill} show the spectrum at 35~km/s and 150~km/s bins, respectively. The on-source spectrum (extracted from an aperture centered on the JWST position) does not show any statistically significant  emission features across the full frequency coverage. An emission feature is seen 0\farcs5 north-east of the JWST position, extended across $\sim 0.4$~arcsec. The tentative line is at 258.7~GHz, implying a spectroscopic redshift of $z = 12.117$ if this is a true \oiiil{} emission line. This spectrum is shown with a 1\,mJy offset for visualization. Note that the larger standard-deviation is caused by the larger aperture used to extract the tentative line. We stress that further observations are necessary to rule out a spurious signal and the association with the target, as discussed in detail in the main text.
{\it Bottom:} The atmospheric transmission at 0.5~mm precipitable water vapour -- similar to the ALMA observing conditions (see Table~\ref{tab:alma_observations}) -- shows only minor absorption features ($< 10$~\%). The four tunings span the redshift range 11.9-13.5, covering 98\% of the confidence limits predicted from multiple photometric redshift methods \citep{Castellano2022,Naidu2022}}
    \label{fig:spectrum_plus_tentative}
\end{figure*}



\bsp	
\label{lastpage}
\end{document}